# Enhancement of superconductivity coexisting with charge density wave in lattice expanded NbTe$_2$

Takaya Shimokawa[1][‡], Yukiko Obata[1], Hayato Makino[1], Kaito Sato[1], Kazutoshi Shimamura[2]
Hiroyuki Okamoto[3], Masao Obata[4, 5], Tatsuki Oda[4, 5], and Yasuo Yoshida[1][*]

[1] *Department of Physics, Kanazawa University, Kanazawa, Ishikawa 920-1192, Japan.*
[2] *Institute of Science and Engineering, Kanazawa University, Kakuma-machi, Kanazawa, 920-1192, Japan.*
[3] *Division of Health Sciences, Kanazawa University, Kakuma-machi, Kanazawa, 920-1192, Japan.*
[4] *Graduate School of Natural Science and Technology, Kanazawa University, Kanazawa 920-1192, Japan*
[5] *Institute of Science and Engineering, Kanazawa University, Kanazawa 920-1192, Japan*

We report a significant enhancement of superconducting transition temperature ($T_c$) of transition metal dichalcogenide (TMD) superconductor NbTe$_2$ from 0.56 K to 2.8 K. Detailed x-ray structure analysis reveals that our $T_c$-enhanced sample has an anisotropic lattice distortion inducing ~1% expansion of the unit cell volume and multi-domain formation in the *ab* planes. Despite the unit cell expansion, the distorted 1T structure, closely related to the charge density wave (CDW) order in this material, persists. Hall measurements show almost identical behaviors for both samples indicating that electronic structure does not change much due to the unit cell expansion. These results suggest that the CDW still coexists with the enhanced superconductivity unlike the other TMD superconductors.

## I. INTRODUCTION

Superconductivity and charge density wave (CDW) orders often coexist in transition metal dichalcogenide (TMD) two-dimensional (2D) superconductors. Opening an energy gap due to CDW formation removes parts of the Fermi surface in 2D systems, lowering the possibility of electron pair formations. This is one of the reasons that TMD superconductors have relatively low superconducting transition temperatures ($T_c$) which are normally below 10 K. By using external tuning parameters like pressure [1,2], intercalation [3–5], electric field [6–8], and reduction of sample dimensions [9,10], CDW tends to be suppressed accompanying with a large enhancement of $T_c$. In most of the cases, the $T_c$s reach the highest when the CDW completely disappears indicating that the superconductivity competes with CDW and the Cooper pairing can be mediated with CDW fluctuations in these superconductors [11,12].

NbTe$_2$ belongs to TMD and has the superconducting transition around 0.5 K [13,14] and CDW transition with 3×1 superstructure at 550 K [15–17]. At the CDW transition, the structure changes from a high symmetry 1T to a distorted 1T structure, so-called 1T'' (C2/m), with double zigzag chains as shown in Fig. 1. Theoretical predictions suggest the existence of topologically nontrivial bands [18], which is confirmed experimentally as a linear magnetoresistance at high magnetic fields [19]. Hydrostatic pressure ~20 GPa induces in NbTe$_2$ the polytype transitions from 1T'' to 1T' (P2$_1$/m) with single zigzag chains, accompanying with the collapse of CDW [20,21]. However, unlike other TMD superconductors, the suppression of CDW does not enhance the $T_c$ of NbTe$_2$ at least down to 2 K. Instead, the superconductivity of NbTe$_2$ is easily suppressed with relatively small pressure ~ 2 GPa [14]. The $T_c$ of NbTe$_2$ drastically increases up to ~ 3 K by Cu intercalation with the expansion of the interlayer distance, maintaining the 1T" structure and CDW [22,23]. These reports are unique among TMD superconductors and may suggest that the superconductivity and CDW are cooperative in NbTe$_2$.

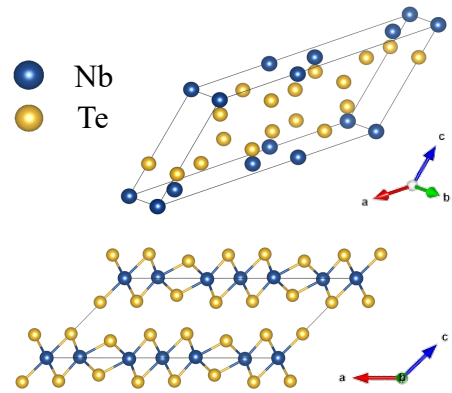

**Fig. 1.** Room-temperature crystal structure of the 1T'' NbTe$_2$ (space group C2/m).

---

[‡]*Email: shimokawa@stu.kanazawa-u.ac.jp*
[*]*Corresponding author: yyoshida@se.kanazawa-u.ac.jp*

Here, we report a large enhancement of $T_c$ from 0.56 K to 2.8 K in NbTe$_2$ coexisting with CDW. X-ray structure analyses show ~ 1% of the unit cell volume expansion in our $T_c$-enhanced sample due to anisotropic distortion of the lattice while maintaining the 1T" structure. We find no significant variation in Hall coefficient $R_H$, also supporting that CDW still coexists with the enhanced superconductivity. Our results suggest that the large enhancement of $T_c$ in NbTe$_2$ differs in the mechanism compared with similar $T_c$ enhancements in other TMD superconductors.

## II. EXPERIMENTAL AND NUMERICAL DETAILS

We grew standard ($T_c$ = 0.56 K) and $T_c$-enhanced ($T_c$ = 2.8 K) NbTe$_2$ single crystals by chemical vapor transport (CVT) method. For standard samples, we first prepared NbTe$_2$ polycrystals with a simple solid-state reaction method by heating as follows. Mixtures of niobium (Nb) powder (99.99 %) and tellurium (Te) powder (99.99 %) were sealed inside a vacuumed (5.0×10$^{-4}$ Pa) quartz ample (16 mm in inner diameter). And then we heated the ample in a muffle furnace at a temperature of 520 °C for 92 h and quenched by furnace cooling to obtain a polycrystalline powder. We heated the polycrystalline powder and iodine (I$_2$) (2 mg/cm$^3$) in a simple tube furnace at reaction end temperature of 920 °C and growth end temperature of 850 °C for 9 days and quenched by furnace cooling. For $T_c$-enhanced samples, we skipped the process of growing polycrystals and grew NbTe$_2$ single crystals by CVT with mixtures of niobium powder, Te powder and I$_2$. Single crystals of standard and $T_c$-enhanced samples were mainly flaky, with a maximum size of about 1 cm. We also used NbTe$_2$ purchased from HQ-graphene as a standard sample. All results of the HQ-graphene sample agree with those of the standard sample we grew.

We characterized crystal structures with out-of-plane X-ray diffraction (XRD) patterns and single-crystal XRD patterns using a Rigaku MiniFlex II equipped with Cu Kα, and Rigaku VariMax RAPID-DW equipped with Mo Kα, respectively. We carried out transport measurements and magnetization measurements with Physical Property Measurement System (PPMS, Quantum Design), and Magnetic Property Measurement System (MPMS, Quantum Design), respectively. We measured resistivity and Hall resistivity using a standard four-probe technique and a six-probe technique, respectively, in magnetic fields applied normal to the 2D layers.

The first-principles electronic structure calculation package, Ecalj [24], was used with the all-electron and mixed basis set methods. The PBE version of the generalized gradient approximation was used. The spin-orbit interaction was considered using the perturbation approach in the density of states (DOS) calculation. The number of k-points was set to 10 × 10 × 6 for the self-consistent field calculation and 30 × 30 × 18 for the DOS calculation. The structural optimization was performed by QUANTUM ESPRESSO [25] within the scalar-relativistic approximation, with cutoff energies of 60 Ryd and 400 Ryd for the wave function and the charge density, respectively.

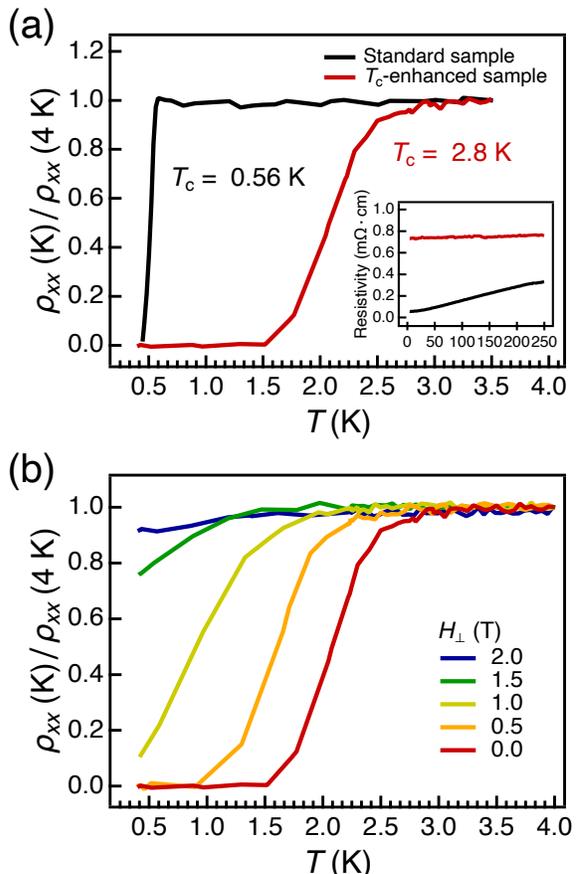

**Fig. 2.** Temperature and magnetic-field dependence of resistivity in NbTe$_2$ samples. (a) Comparison of superconductivity transition temperature ($T_c$) in standard samples (black line) and $T_c$-enhanced samples (red line). The inset shows the resistivities of the samples at higher temperatures. (b) The critical temperatures of the $T_c$-enhanced in magnetic fields.

## III. RESULTS AND DISCUSSION

Fig. 2(a) shows the temperature dependence of resistivity of the standard and $T_c$-enhanced samples obtained with two different growth procedures. The $T_c$-enhanced sample has the onset temperature at 2.8 K which is 5 times higher than that of the standard sample ($T_c$ = 0.56 K). To further characterize the $T_c$-enhanced sample, we estimate the critical field $H_{c2}$ by measuring the critical temperatures in magnetic fields as shown in Fig. 2(b). We estimate the critical magnetic field ($H_{c2}(0)$) based on Ginzburg Landau expression:

$$H_{c2}(T) = H_{c2}(0) \frac{[1-(T/T_c)^2]}{[1+(T/T_c)^2]}$$

The $T_c$-enhanced sample shows a vast enhancement of $H_{c2}(0) \sim 2.5$ T compared to that of 0.04 T of the standard one. We also estimate the residual resistance ratio (RRR) of the samples based on the higher temperature resistivities shown in the inset of Fig. 2(a). The RRR of the $T_c$-enhanced sample is as small as ~1 while the RRR of the standard sample is 5-6 which is identical to the values reported in the previous works [13,14,19], obviously suggesting that the $T_c$-enhanced sample contains more disorders than the standard sample does.

To confirm the superconductivity of the $T_c$-enhanced sample, we performed magnetization measurements. Figure 3 shows that the temperature dependence of the magnetization of the $T_c$-enhanced sample. The magnetization in zero-field cooling (ZFC) starts deviating from the one in field cooling (FC) at ~ 2.5 K clearly indicating that the shielding effect. We estimate the superconducting volume fraction to be 0.025%. Since the superconducting transition observed in the resistivity measurements is not quite sharp for the $T_c$-enhanced sample, we probably need to measure the magnetization at lower temperatures to evaluate the volume fraction more precisely.

To understand the difference in $T_c$ between the samples, we first performed out-of-plane XRD measurements. Figure 4(a) shows the XRD patterns on the (00$l$) of $T_c$-enhanced and standard samples. The clear peaks for the standard sample indicate a single phase. The clear main peaks with the tiny sub peaks at higher angles for the $T_c$-enhanced sample indicate that the crystal is nearly a single phase but contains small lattice constant fluctuations, which is consistent with the Laue analysis and magnetoresistance measurements described later. The peaks of the standard sample are consistent with previous studies, but those of the $T_c$- enhanced sample shift to a lower angle [19,22,23]. The interlayer distances $d$ of the standard and $T_c$-enhanced samples are estimated to be 6.628 Å and 6.687 Å, respectively. The $T_c$-enhanced sample expands in the (00$l$) direction by ~ 0.06 Å, which is similar order to the values reported in intercalated and doped TMD material [3,4,22].

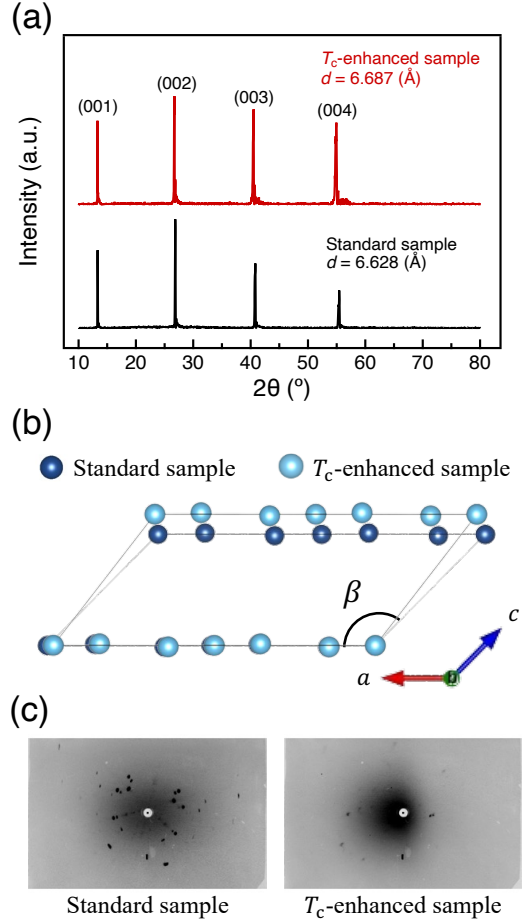

Fig. 4. (a) The out-of-plane XRD patterns of the standard (black line) and $T_c$-enhanced (red line) samples at room temperatures. The $d$ is interlayer distance. (b) Schematic of the observed structural modification in NbTe$_2$. The blue atom and the sky blue atom show Nb of the standard and the $T_c$-enhanced samples, respectively. (c) Laue diffraction patterns of the standard and $T_c$-enhanced samples.

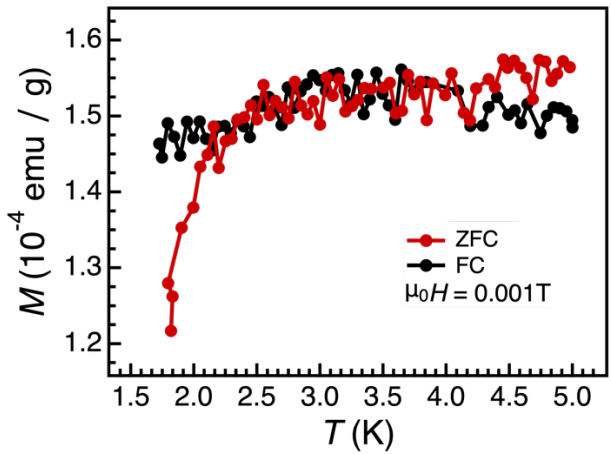

Fig. 3. Temperature dependence of magnetization in zero field cooling (ZFC) and field cooling (FC).

**Table 1.** Lattice parameters and unit cell volume of the standard and the $T_c$-enhanced NbTe$_2$ crystals determined with single-crystal XRD at room temperatures.

| Parameter | Standard sample | $T_c$-enhanced sample | Reference [24] |
|---|---|---|---|
| $a$ (Å) | 19.436(3) | 19.276(6) | 19.392 |
| $b$ (Å) | 3.6451(5) | 3.6732(8) | 3.6420 |
| $c$ (Å) | 9.327(2) | 9.313(2) | 9.375 |
| $\beta$ (°) | 134.58(1) | 133.89(2) | 134.57 |
| Interlayer distance (Å) | 6.643(2) | 6.711(3) | 6.678 |
| Unit cell volume (Å$^3$) | 470.6(2) | 475.1(3) | 471.7 |

We performed single-crystal XRD analysis for further structural characterization as summarized in Table. 1. The grain size used for the measurement is order of ~ 10 μm. As schematically shown in Fig. 4(b), the $T_c$-enhanced sample has a shorter lattice constant along the $a$-axis and a smaller $β$ angle, resulting in a unit cell volume approximately 1% larger than the standard sample [26]. The increase in the interlayer distance is consistent with that obtained from the out-of-plane measurements. Regardless of this unit cell volume expansion, the crystal structure preserves the same space group strongly suggesting that the $T_c$-enhanced sample has the CDW order below room temperatures.

The Laue analysis clarifies the multi-domain formation on the $ab$-plane in the $T_c$-enhanced sample. The Laue diffraction pattern of the $T_c$-enhanced sample shows much fewer spots compared to that of the standard (Fig. 4(c)). Multiple domains in the $ab$-plane should exist also in the standard sample as observed microscopically in the previous work using a scanning tunneling microscope (STM) as a consequence of the 1T-1T" structure transition [17]. This is like in the case of iron-based superconductors, in which many domains with twin boundaries are induced by the lattice deformation due to the structure phase transition from the tetragonal to the orthorhombic lattices [27]. For the $T_c$-enhanced sample, this effect is probably more enhanced due to the significant change in $β$ angle leading to the further reduction in the domain size.

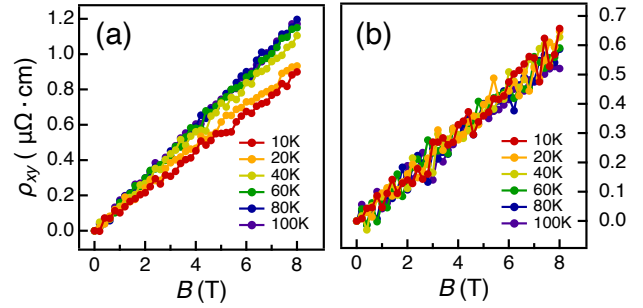

**Fig. 6.** Field dependence of Hall resistivity $ρ_{xy}$ up to 8 T at various temperatures for (a) standard and (b) $T_c$-enhanced NbTe$_2$ crystals.

This reduction in the domain size shortens the mean free path and reduces RRR. This also affects the MR. Based on the standard two carrier model for MR, if the relaxation times $τ$s of the carriers decrease drastically due to multidomain formations or impurities, MR also will be reduced significantly [28]. Figures 5 shows magnetoresistance (MR) defined as $(ρ_{xx}(B,T) - ρ_{xx}(0,T)) / ρ_{xx}(0,T)$ of the samples at various temperatures in fields up to 8 T. MR of the standard sample reaches ~ 5% which is comparable to the previous report [19], while strongly suppressed in the $T_c$-enhanced sample. In other TMD materials, multi-domain formation and grain size reduction lead the suppression of CDW [9,10,29]. At the same time, a strong $T_c$ enhancement or induced superconductivity occur in most of the cases. However, in the $T_c$-enhanced NbTe$_2$ sample, CDW persists as we see from the structure analysis.

To better understand the differences in the electronic structures of the two samples, we performed Hall resistivity measurements. The Hall coefficient $R_H$ is defined as the magnetic field derivative of hall resistivity. As shown in Fig. 6, both the standard and $T_c$-enhanced samples show a positive $R_H$, positive linear dependence of the Hall resistivity on the magnetic field, indicating that holes are the dominant carriers in these samples. A weak temperature dependence of the $R_H$ exists for the standard sample while it is absent for the $T_c$-enhanced sample. The carrier density ~ $1/R_H$ for the $T_c$-enhanced sample slightly increases compared with the standard sample. But other than those small changes, there is no significant difference between the two samples. Sign reversal of $R_H$ associated with CDW transitions and loss of sign reversal of $R_H$ associated with suppression of CDW by external perturbations have been reported in TMD materials. [30–32]. The sign reversal occurs due to the gap opening or closing due to CDW formation or collapsing. The gap opening (closing) removes (increases) the carrier in the part of the Fermi surface and changes the dominant carrier. Such $R_H$ sign reversal is also reported for in NbTe$_2$ at the external pressure where CDW disappears [21]. If CDW transition is suppressed in the $T_c$-enhanced sample, the sign of the $R_H$ can be negative, but that is not the case here. This also suggests that the CDW still coexists with the superconductivity in the $T_c$-enhanced sample. A sign reversal of $R_H$ is also absent for copper-doped NbTe$_2$ in which the $T_c$ is strongly enhanced [22]. We do not know exactly the CDW transition temperature of the $T_c$-enhanced sample, but at least, we don't see any clear anomaly in the resistivity related to CDW transition up to 300 K as in the case for copper doped NbTe$_2$ [22,23].

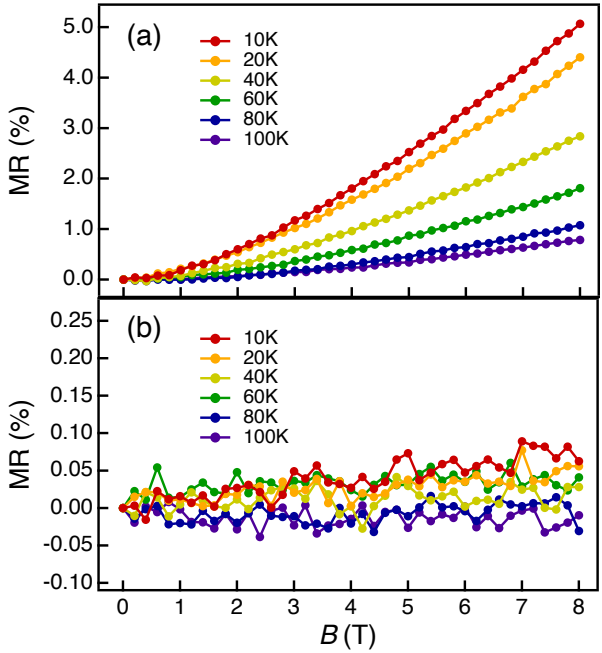

**Fig. 5.** Magnetoresistance (MR) of NbTe$_2$ samples at different temperatures. (a) Standard sample. (b) $T_c$-enhanced sample.

To investigate possible origins for the $T_c$-enhancement, we performed first principles calculations for both samples based on the lattice parameters determined experimentally in Table 1. According to BCS theory, the DOSs at $E_F$ increases, the more $T_c$ improves. This could happen if the lattice parameter changes. However, as we can see in Fig. 7, the DOS at the Fermi level $E_F$ ($E = 0$) has no difference between two samples. The expansion of the unit cell volume softens the lattice leading a stronger electron-phonon coupling $\lambda$, which also improves $T_c$. To check this possibility, it requires more calculations or thermodynamic measurements in future works.

In the monolayer limit of NbTe$_2$ [33], various CDW phases appear at different growth temperatures. A peculiar long periodic CDW ($\sqrt{19} \times \sqrt{19}$) called David star is stabilized at higher temperatures, which is discussed related to the Mott phases [34]. Although no experimental reports show the relationship between such CDW and superconductivity, it is well-known that the Mott phase is closely related to high-$T_c$ cuprates [35]. As described before, we used basically the same method to synthesize NbTe$_2$ crystals, but we skipped one process for the $T_c$-enhanced sample. A possible reason for the difference between the two samples is the difference in temperature gradient during crystal growth. If the growth temperatures are slightly different, different types of CDWs can exist in the samples. Hence it will be very informative if we observe which types of CDWs are stable in our $T_c$-enhanced sample and the relationship with the spatial distribution of the superconducting gaps. For further studies, a microscopic investigation using an STM will be a key to unveil the mechanism of the observed $T_c$ enhancement, but it is beyond the scope of this paper.

## IV. CONCLUSION

In conclusion, we find that NbTe$_2$ with the unit cell expansion (~ 1%) has 5 times higher $T_c$ compared to that of the standard sample. We conclude that the enhanced superconductivity and CDW order coexist in the $T_c$-enhanced sample based on the results of the structure analysis and transport and magnetic measurements unlike the $T_c$ enhancements observed in other TMD superconductors. The first principles calculations show that the unit cell volume expansion does not give the enhancement of DOS at $E_F$, excluding that as a possible origin of the observed $T_c$ enhancement. For a deeper understanding, direct observation of the CDW and superconducting gap using STM is essential.

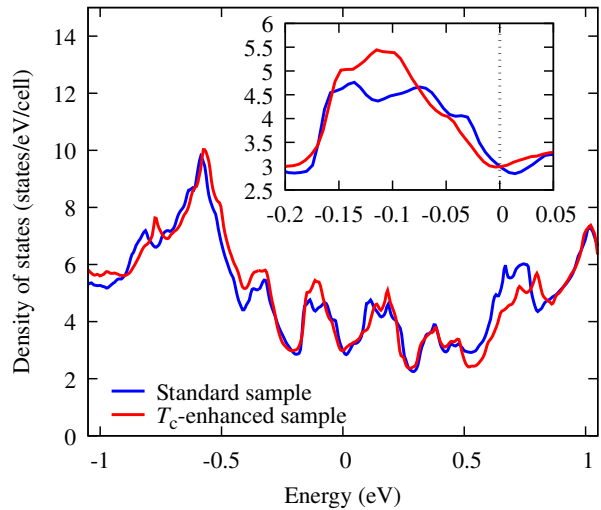

**Fig. 7.** The density of states calculations of NbTe$_2$ based on DFT theory with spin-orbit interaction. Red and blue lines indicate the results for $T_c$-enhanced and standard samples. The lattice parameters experimentally obtained are used in the calculation. The result near $E_F$ ($E = 0$) is enlarged in the inset.

## ACKNOWLEDGMENTS

We thank Koichiro Ienega for fruitful discussions. We also thank Yuji Matsumoto for technical helps for X-ray analyses. The computation in this work was performed using the facilities of the Supercomputer Center, Institute for Solid State Physics, The University of Tokyo, Japan. This work was supported by The Mitani Foundation for Research and Development, and a JSPS Grant-in-Aid for Scientific Research (KAKENHI) Grant Number 23K26541, 23H05459, 24K06955.